\documentclass[conference]{IEEEtran}
\IEEEoverridecommandlockouts
\usepackage{cite}
\usepackage{amsmath,amssymb,amsfonts}
\usepackage{algorithmic}
\usepackage{graphicx}
\usepackage{textcomp}
\usepackage{xcolor}
\usepackage{hyperref}
\usepackage{multirow}

\def\BibTeX{{\rm B\kern-.05em{\sc i\kern-.025em b}\kern-.08em
    T\kern-.1667em\lower.7ex\hbox{E}\kern-.125emX}}

\begin{document}

\title{Exploring Emotions in EEG: Deep Learning Approach with Feature Fusion
}

\author{\IEEEauthorblockN{1\textsuperscript{st} Danastan Tasaouf Mridula}
\IEEEauthorblockA{\textit{dept. of CSE} \\
\textit{University of Dhaka}\\
Dhaka, Bangladesh \\
danastantasaouf-2020519542@cs.du.ac.bd}
\and
\IEEEauthorblockN{2\textsuperscript{nd} Abu Ahmed Ferdaus*}
\IEEEauthorblockA{\textit{dept. of CSE} \\
\textit{University of Dhaka}\\
Dhaka, Bangladesh\\
ferdaus@du.ac.bd} 
\and
\IEEEauthorblockN{3\textsuperscript{rd} Tanmoy Sarkar Pias*}
\IEEEauthorblockA{\textit{dept. of CSE} \\
\textit{Virginia Tech, USA}\\
Blacksburg, USA \\
tanmoysarkar@vt.edu}
\and

}

\maketitle

\begin{abstract}
Emotion is an intricate physiological response that plays a crucial role in how we respond and cooperate with others in our daily affairs. Numerous experiments have been evolved to recognize emotion, however still require exploration to intensify the performance. To enhance the performance of effective emotion recognition, this study proposes a subject-dependent robust end-to-end emotion recognition system based on a 1D convolutional neural network (1D-CNN). We evaluate the SJTU\footnote{\href{https://en.wikipedia.org/wiki/Shanghai_Jiao_Tong_University}{Shanghai Jiao Tong University(SJTU)}} Emotion EEG Dataset SEED-V with five emotions (happy, sad, neural, fear, and disgust). To begin with, we utilize the Fast Fourier Transform (FFT) to decompose the raw EEG signals into six frequency bands and extract the power spectrum feature from the frequency bands. After that, we combine the extracted power spectrum feature with eye movement and differential entropy (DE) features. Finally, for classification, we apply the combined data to our proposed system. Consequently, it attains 99.80\% accuracy which surpasses each prior state-of-the-art system.
\end{abstract}


\section{Introduction}
The concept of emotion detection seems an alluring research topic for humankind. Emotion is associated with both our physical health and mental health \cite{peng2022joint}. For instance, sometimes we feel intense chest pain when we are emotionally sad. Now, researchers conduct emotion recognition experiments in terms of different aspects including computer science \cite{9751142}, neuroscience \cite{krumhuber2023role}, psychology, medical diagnostics, and cognitive science.
The study of emotion recognition not only recognizes mental state but also assists in recovering from anxiety \cite{minkowski2021feature}, depression, neurological disorders, and driving fatigue \cite{fang2022classification}. The emotion recognition experiment has two categories from the viewpoint of data sources: non-physiological signals-based and physiological signals-based. Considering non-physiological signals, for instance, facial expression \cite{saganowski2022emognition}, voice intonation, and gestures: the emotion recognition system is not effective as people can conceal emotions. Besides, the people who are physically disabled and can not speak, it is pretty tough to realize their emotions through their gestures and activities. 
Consequently, investigators utilize physiological signals for emotion classification. C. -J. Yang \cite{yang2020convolution} aimed to generate an emotion recognition system based on the electrocardiogram (ECG) and photoplethysmography (PPG) signals for three emotional states including positive, negative, and neutral. Mahrukh  \cite{mahrukh2023sentiments} proposed an automatic labels-generating approach from movie subtitles and used brain fMRI images for positive, negative, and neural sentiment analysis.
Among several physiological signals like electrocardiogram (ECG), functional magnetic resonance imaging (fMRI), and electroencephalogram (EEG): EEG reveals promising performance in real-time emotion recognition. Recently, with the improvement of brain-computer interaction \cite{hinss2023open} and the use of EEG signals, it seems human emotion detection systems are reliable. However, a single modality is not fully appropriate for emotion recognition as the different modality elicits different prospect and beneficial information. In \cite{zhang2022deep}, Zhang proposed a Deep Emotional Arousal Network (DEAN) for multimodal sentiment analysis and emotion recognition. Though immense emotion recognition research has been done using electroencephalogram(EEG) signals or multi-modalities fusion techniques but still requires more scrutinize.
Our main contributions to this study are as follows:
\begin{itemize}
    \item We combine the power spectrum with eye movement and differential entropy (DE) features.
\end{itemize}
\begin{itemize}
    \item We propose a robust subject-dependent end-to-end emotion recognition system applying 1D-CNN. Also, we evaluate our method on the SEED-V dataset which achieves 99.80\% accuracy.
\end{itemize}
\section{Literature review}
\begin{table*}
    \centering
    \caption{LITERATURE REVIEW ON SEED-V Dataset}
    \label{tab:one}
    \begin{tabular}{|c|c|c|c|}
         \hline
         \textbf{Year} & \textbf{Research} & \textbf{Features} & \textbf{Related ML/DL algorithms} \\ \hline
         2021 & Wei Liu \cite{liu2021comparing} & EEG, Eye Movement & Bimodal deep autoencoder (BDAE) \\ \hline
         2020 & Xun Wu\cite{wu2022investigating} & EEG,Eye Movement & Deep Canonical Correlation Analysis \\ \hline
         2020 & Yu-Ting Lan1\cite{lan2020multimodal} & EEG, EIG, and Eye Movement & Deep Generalized Canonical Correlation Analysis \\ \hline
         2019 & Luo\cite{luo2019gan} & Differential Entropy, Eye Movement & Conditional Boundary Equilibrium GAN (cBEGAN) \\ \hline
         2019 & Jiang-Jian Guo \cite{guo2019multimodal} & EEG, Eye Movement, Eye Image & Hybrid CNN-LSTM \\ \hline
         2019 & Tian-Hao Li\cite{li2019classification} & EEG, Eye Movement & SVM \\ \hline
    \end{tabular}
\end{table*}

\begin{figure*}
    \includegraphics[height=7.3cm]{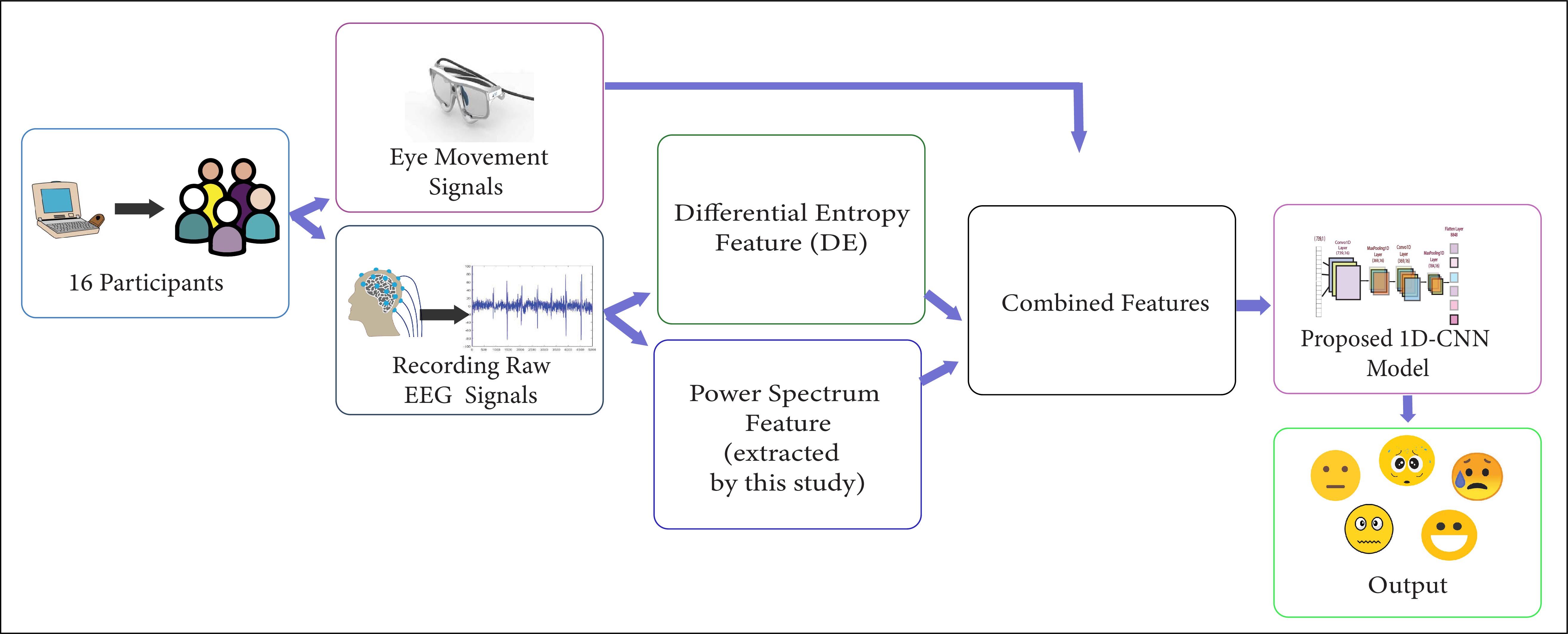}
    \caption{Complete workflow of the proposed system on the SEED-V dataset.
}
    \label{fig:pipeline}
\end{figure*}

Researchers have delved into a significant amount of experiments based on EEG signals using various publicly available datasets, including DEAP \cite{akter2023evaluating}, SEED \cite{wang2023self}, LUMED-2 \cite{alam2023human}, DREAMER \cite{liu2021comparing}, and MAHNOB-HCI \cite{pan2023multimodal} and feature extraction methods up to now. 
These days, machine learning-based emotion recognition approaches demonstrate better performance. Ahmed \cite{ahmed2023emotion} applied different ML techniques, including SVM (support vector machine), and KNN (k-nearest Neighbour) for emotion recognition. K. S. Kamble \cite{kamble2021ensemble} compared five ensembles learning-based ML approaches with five conventional ML approaches to classify emotions. However, ML algorithms elicit poor results in extracting deep features and nonlinear patterns \cite{zhang2020emotion} whilst deep learning techniques are convenient for extracting complex features \cite{samavat2022deep}.  In \cite{hassan2020human}, R. Hassan proposed a hybrid model with CNN and LSTM to detect real-time emotion recognition.
In comparison with other deep learning models, a convolutional neural network (CNN) is a fruitful neural network model with the ability to extract high-layer semantic features. These days, CNN is widely used in signal processing \cite{9564204}, image classification \cite{barman2022deep} etc. The research community acclaims that CNN manifests good performance in emotion analysis. In \cite{liu2020subject}, Liu proposed a subject-independent algorithm, a dynamic empirical convolutional neural network (DECNN) for emotion recognition. In another study,  \cite{akter2022m1m2} Akter used two different convolutional neural network (CNN) models for two levels of valence, and arousal utilizing the DEAP emotion recognition dataset. Nevertheless, numerous studies have been done with SEED and SEED-IV \cite{jiang2023emotion,feng2022eeg} for neural pattern analysis and emotion recognition. A. Samavat \cite{samavat2022deep} claimed the gamma and beta band was beneficial for EEG-based emotion recognition and  N. Priyadarshini \cite{priyadarshini2023emotion} initiated a multimodal fusion network to find out better discriminative information using Gated Recurrent Unit (GRU) and Long-Short Term Memory (LSTM). Y. Jiang \cite{jiang2023emotion} brought forward a novel attention mechanism-based multiscale feature fusion network (AM-MSFFN) that could extract high-level features at different scales. Furthermore, L. Feng \cite{feng2022eeg} was captivated to analyze the topological information of each brain segment applying ST-GCLSTM. Among three SEED emotion dataset versions including SEED, SEED-IV, and SEED-V, this study has investigated SEED-V. To the best of our knowledge, a handful of research has been done on the SEED-V emotion dataset. Table \ref{tab:one} illustrates some prior research regarding the SEED-V emotion dataset.


\section{Methodology}

In this study, we propose a subject-dependent end-to-end emotion detection system. A complete workflow of the proposed system is demonstrated in Fig. \ref{fig:pipeline}. At first, we explore the SJTU SEED-V emotion dataset. After that, we decompose raw EEG signals and extract the power spectrum feature from sub-bands. Afterward, we combine all features and fit into a 1D-CNN model. Finally, the model recognizes emotion from it.
\subsection{The SEED-V Dataset
}

\begin{figure}
    \includegraphics[width=0.5\textwidth]{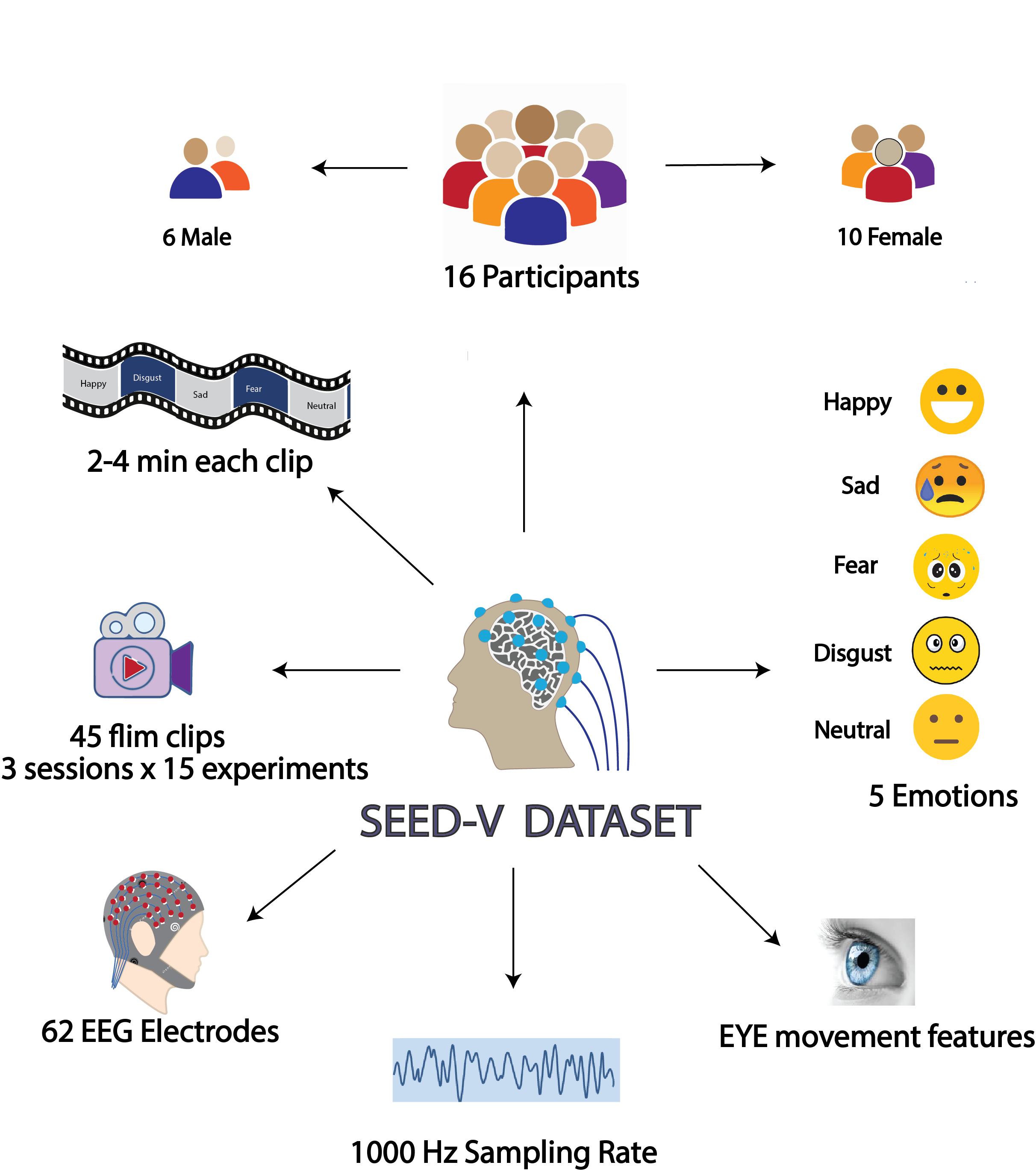}
    \caption{Overview of SEED-V Dataset.}
    \label{fig:seed-v}
\end{figure}

BCMI laboratory provides the SJTU Emotion EEG Dataset (SEED-V). This dataset is freely available to the academic community as a modified version of the SEED dataset. This dataset comprises EEG signals and eye movement signals of 16 participants for five emotions (happy, fear, sad, disgust, and neutral). Besides, their team provides an extracted differential entropy(DE) feature. An overview of the SEED-V dataset is shown in Fig. \ref{fig:seed-v}.

\begin{figure}
    \centering
    \includegraphics[width=0.15\textwidth]{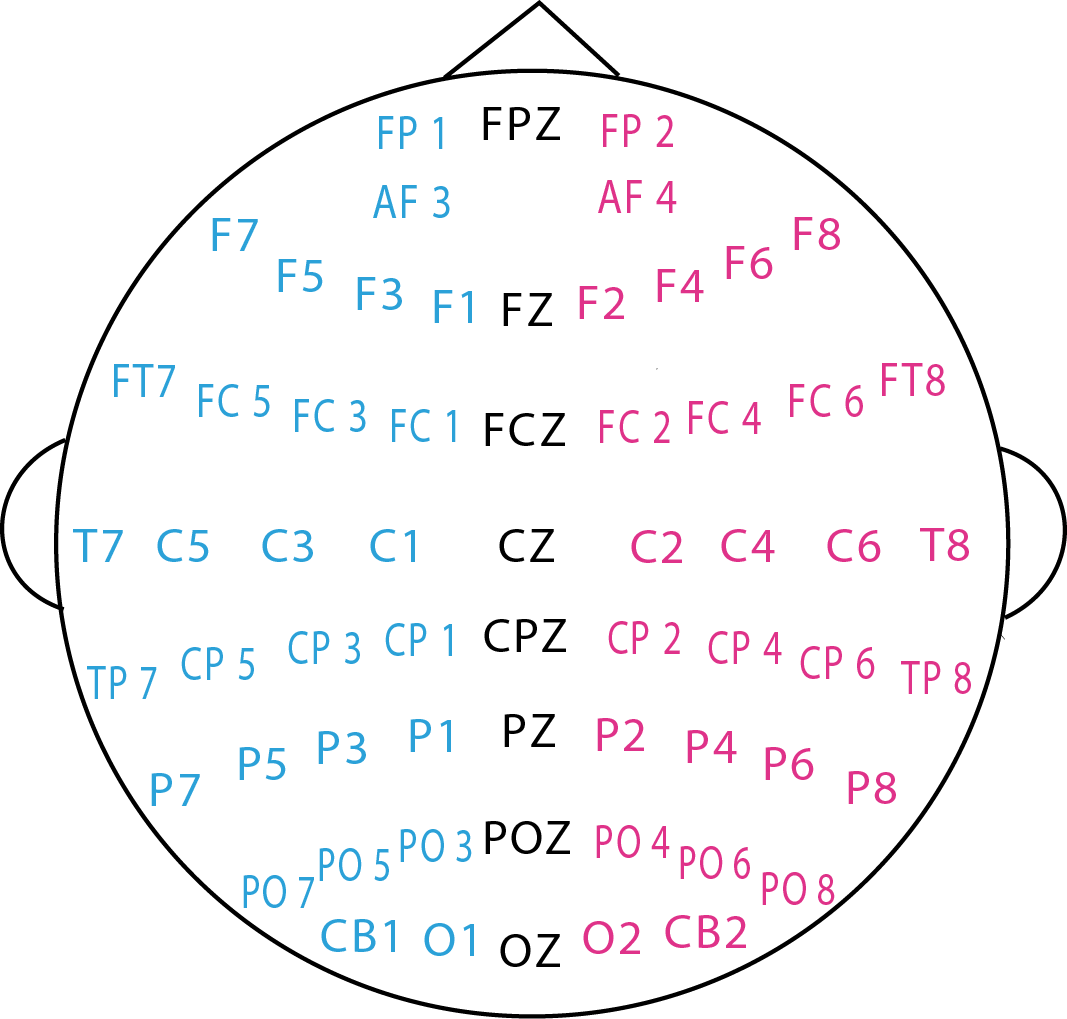}
    \caption{62-channel ESI NeuroScan System}
    \label{fig:electrode}
\end{figure}
The investigator recorded EEG signals through a 62-channel ESI NeuroScan System \ref{fig:electrode} and eye movement signals with SMI eye-tracking glasses. Each participant experienced 3 sessions where they watched 15 film clips in each session. Hence, the total number of trials for 3 sessions is 3*15 = 45. Finally, the total number of trials is 45*16 = 720 for 16 participants. They exhibited 3 film clips per emotion in one session and the same emotion clip did not play consecutively. Each clip duration was 2-4 minutes with distinct content. Table \ref{tab:two} illustrates the characteristics of the SEED-V dataset. Participants provided their ratings of sad, happy, disgust, neural, and fear according to their point of view.  
    
\subsection{Preprocessing}
EEG signals make noise instinctively due to artifacts, for instance, eye blink, saccade, muscle movement, and head movement. Such raw data can interrupt while analysis. However, from the raw EEG signals, the SEED-V dataset team has extracted the differential entropy feature. Consequently, the pre-processed differential entropy (DE) feature is unchained from any noise and convenient for analysis.
\subsection{Feature extraction}
In comparison with other feature extraction techniques, FFT has shown fruitful results and helps to speed up the training process. In this study, among various feature extraction techniques, we utilize the Fast Fourier Transform (FFT).
\subsubsection{Power Spectrum feature extraction}
We decompose raw EEG signals into six frequency bands (0.5,4,7,12,16,30,100) with FFT. After that, we extract the power spectrum feature using the bin\_power function from the PyEEG module of Python from the bands with 66 channels. The power spectrum denotes how much power is contained in the frequency components of the signal.\footnote{\href{https://www.wikilectures.eu/w/Frequency_and_power_spectra}{https\:\/\/www\.wikilectures\.eu\/w\/Frequency\_and\_power\_spectra}}
Equation \ref{one} depicts the power spectrum  as follows:
\begin{equation}
\label{one}
S_{xx}(f) = \int_{-\infty}^{\infty}R_{xx}(\tau)e^{-i2\pi f\tau}d(\tau)
\end{equation}
where $R_{xx}(\tau)$ denotes autocorrelation function of the signal.
The power of the signal for a given frequency band is [f1,f2], where  \(0 < f1 <f2 \). 
\begin{equation}
\label{two}
    P_{bandlimited}=2\int_{f1}^{f2}S_{xx}(f)df
\end{equation}
As $S_{xx}(-f) = S_{xx}(f)$, from a positive and negative frequency band, a similar amount of power can be attributed, which is depicted as equation \ref{two} 
\subsubsection{Differential Entropy feature extraction}
SEED-V team has provided a pre-processed noise-free differential entropy(DE) feature. First of all, with a 200Hz sampling rate, the raw EEG data are downsampled. Afterwards, via a bandpass filter between 1Hz and 75Hz, the EEG data are processed. 62 channels along with 5 sub-bands are used to extract the DE feature.
\begin{table}[]
\centering
\caption{No of participants, no of emotions, no of sessions, no of trials, and movie clip duration for the SEED-V dataset.}
\label{tab:two}
\begin{tabular}{|p{2.5cm}|p{2cm}|}
    \hline       
    \textbf{No of  Participants} & 16 \\ \hline
   \textbf{No of Emotions} & 5  \\ \hline
    \textbf{No of sessions} & 3 \\ \hline
    \textbf{No of Trials} & 720 \\ \hline
    \textbf{Movie clip duration} & 2-4 minutes \\ \hline
\end{tabular}
\end{table}

\subsubsection{Eye Movement feature extraction}
Several features from various parameters, including pupil diameter, fixation, dispersion, saccade, blink, and event statistics are extracted as eye movement features through SMI ETG eye-tracking glasses. Finally, as eye movement features, a total number of 33 features are selected. 
         

\begin{table*}[]
    \centering
    \caption{ Data array and Label array’s Shape of Power Spectrum, Differential Entropy, Eye Movement and Combined data}
    \label{tab:six}  
    \begin{tabular}{|c|p{2.5cm}|p{11cm}|}
    \hline
    \textbf{Feature} &  \textbf{Shape} &  \textbf{Contents}\\ \hline
        \multirow{2}{2cm}{Power Spectrum} &  Data:(2352720, 396) & Data: (rows = windows , cols = 66 channels x 6 bands)  \\
        & Label:(2352720, 5) & Label: (rows = windows, cols = 5 emotions) \newline \\ \hline
        \multirow{2}{2cm}{Differential Entropy} & Data: (29168, 310) & Data: (rows = 3 sessions data x 16 participants, cols = 62 channels x 5 bands) \\
        & Label: (29168, 5) & Label: (rows = 3 sessions data x 16 participants, cols = 5 emotions) \newline \\ \hline
        \multirow{2}{2cm}{Eye Movement} & Data: (29168, 33) & Data:(rows = 3 sessions data 
 x 16 participants, cols = 33 features) \\
        & Label: (29168, 5) & Label: (rows = 3 sessions data 
 x 16 participants, cols = 5 emotions) \newline \\ \hline
        \multirow{2}{2cm}{Combined Data} & Data: (2352720, 739) & Data: rows=windows, cols = (66 channels x 6 bands)+(62 channels x 5 bands)+(33 eye features) \\
        & Label:(2352720, 5) & Label: (rows = windows, cols= 5 emotions) \\ \hline
      
    \end{tabular}
\end{table*}
\subsection{Data Preparation}
In this study, we combine eye movement, DE, and power spectrum features. All features data array and label array shape and contents are listed in Table \ref{tab:six}. After applying FFT, we obtained  2,352,720 rows from 720 raw EEG files and 396 columns for the power spectrum feature. The columns denote the result of 66 channels and 6 sub-bands. The rows denote window slices. Besides, the label array consists of 2,352,720 rows and 5 columns. Eye movement features data array rows denote the merge eye movement data of 16 participants. One participant's eye movement data array is (1823,33). Therefore, 16 participants merged data array size is (1823 x 16, 33)=(29168,33). Likewise, the row number of DE's data array denotes the merged DE data of 16 participants whereas the column number demonstrates 5 bands and 62 channels.

       
Prior to combining data, we are required to do padding in the eye movement and DE feature. For clarification, we demonstrate an example of the process of combined data. To begin with, we pass 1\textunderscore1\textunderscore1\textunderscore20180804.cnt.npy (1st EEG raw file) into our declared FFT function whose shape is (66, 72000). After FFT,data shape becomes (1437,396) and saves in 1\textunderscore1\textunderscore1\textunderscore\text{FFT.npy} file. The corresponding eye movement and DE file’s (1\textunderscore1\textunderscore1\textunderscore{EYE.npy}), (1\textunderscore1\textunderscore1\textunderscore{DE.npy}) shape is (18,33) and (18,310). So, we do padding using repeat and insert function from the python numpy library to (1\textunderscore1\textunderscore1\textunderscore{EYE.npy}) and (1\textunderscore1\textunderscore1\textunderscore{DE.npy}) for increasing row numbers like 1\textunderscore1\textunderscore1\textunderscore\text{FFT.npy} file. After padding, (1\textunderscore1\textunderscore1\textunderscore{EYE.npy}),(1\textunderscore1\textunderscore1\textunderscore{DE.npy}) files shape are (1437,33) and (1437,310) respectively. Finally, we stack the 3 files including 1\textunderscore1\textunderscore1\textunderscore\text{FFT.npy}, (1\textunderscore1\textunderscore1\textunderscore{EYE.npy}), (1\textunderscore1\textunderscore1\textunderscore{DE.npy}) and shape becomes (1437,739). We follow the same procedure for each file and finally stack all the files.

\subsection{Model Architecture}
\begin{figure*}
    \centering
    \includegraphics[width=0.75\textwidth]{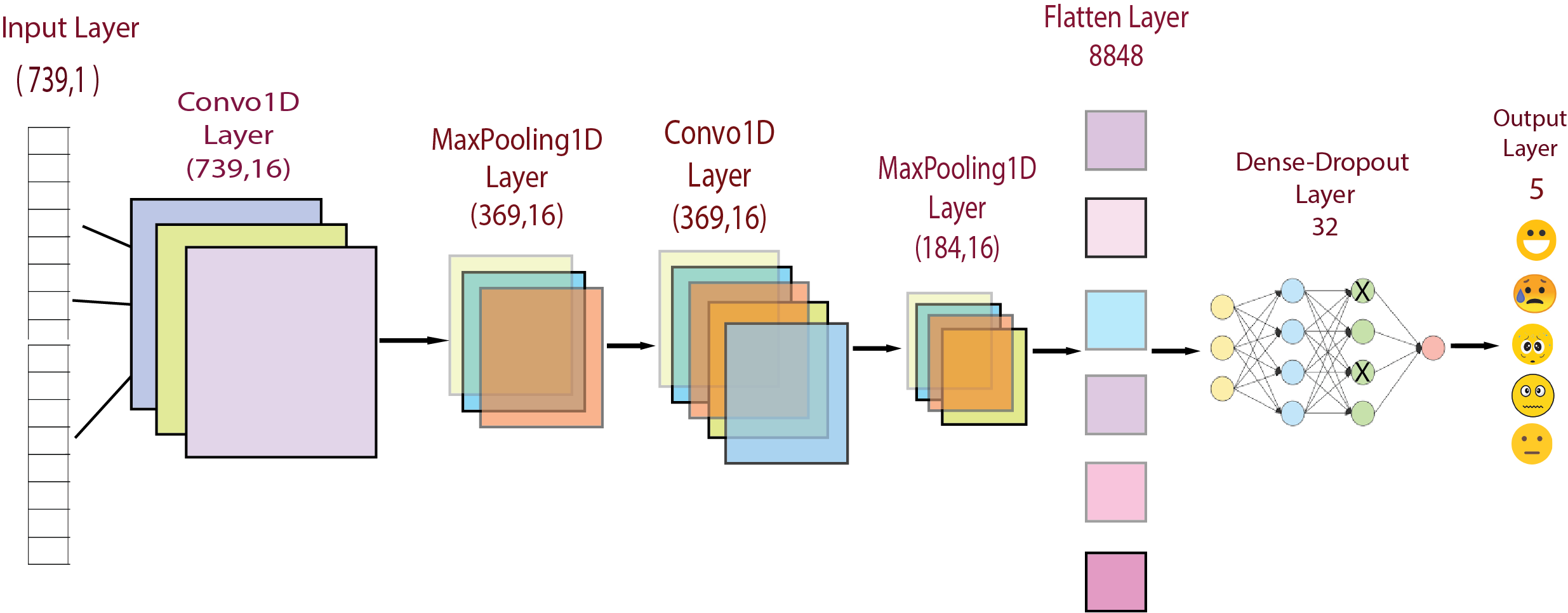}
    \caption{Our proposed 1D-CNN model architecture}
    \label{fig:cnn}
\end{figure*}

In the Literature review section, we have elaborately discussed various deep learning algorithms for signal recognition that are used by many researchers. Among those, ANN \cite{pias-vehicle-ANN, pias-gender-ANN}, CNN \cite{pias-ECG-CNN, pias-vehicle2-CNN, pias-EEG-CNN}, hybrid CNN-LSTM, LSTM, BDAE, and DCCA are frequently applied to the SEED-V dataset. However, in this study, we utilize a 1D-CNN model with two convolution layers. 1D-CNN has vast advantages including the ability to recognize complex features, parameter sharing, and the capability of extracting local patterns from sequential data. Our proposed 1D-CNN architecture is demonstrated in Fig. \ref{fig:cnn}. This proposed method comprises two convolution layers, two max-pooling layers, and one output layer after the dense layer.
\section{Experiment and Result}
\begin{figure}
    \centering
    \includegraphics[width=0.5\textwidth]{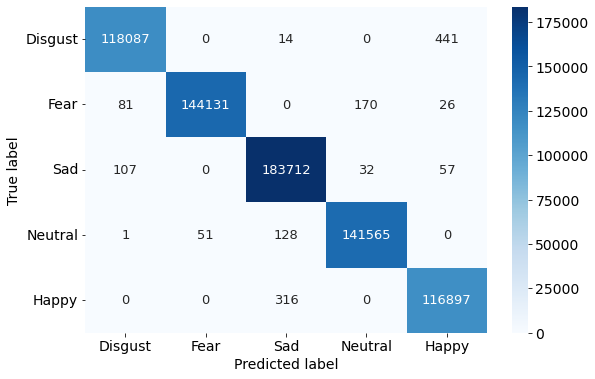}
    \caption{Confusion matrix}
    \label{fig:cm}
\end{figure}

The data pre-processing, feature extraction, and experiment with a single modality including eye movement, DE, and power spectrum features have been done in Google Colab with the Keras library. Later, we do our experiment on a Linux Virtual Machine with combined data. We use both Keras and tensorflow version 2.6.
\begin{table}[]
    \centering
    \caption{Size of Train Data \& Label, Test Data \& Label}
    \label{tab:nine}
    \begin{tabular}{|c|c|}
    \hline
         \textbf{Array} & \textbf{Shape}  \\ \hline
        Train Data & (1646904, 739) \\ \hline
        Train Label & (1646904, 5) \\ \hline
        Test Data & (705816, 739) \\ \hline
        Test Label & (705816, 5) \\ \hline
    \end{tabular}
\end{table}
We utilize FFT as a feature extraction technique. We divide 70 percent data as training and the rest 30 percent of data as testing. The array size of the train and test is indicated in Table \ref{tab:nine}. We normalize Train and Test data using the StandardScaler() function from the Python sklearn library. Apart from this, the optimizer and the loss function of the proposed method were Adam, categorical\_crossentropy respectively. Besides, batch size is declared at 1000  as a hyperparameter. We classify five emotional states by applying the proposed model. And our proposed method obtains 99.80\% test accuracy within 2 milliseconds.  
\begin{table*}[]
    \caption{Single modality VS multi-modality performance on SEED-V dataset}
    \label{tab:ten}
    \begin{center}

    \begin{tabular}{|p{6cm}|p{7cm}|p{2.4cm}|}
    \hline
        \textbf{Model} & \textbf{Features} & \textbf{Testing Accuracy}\\ \hline
        A simple neural network with 3 hidden layers & Differential Entropy & 26\% \\ \hline
        A simple neural network with 3 hidden layers & Eye Movement & 61\% \\ \hline
        Proposed model & Power Spectrum & 73\% \\ \hline
        Proposed model & Combined features (Power spectrum, Eye movement, Differential Entropy) & 99.80\% \\ \hline
    \end{tabular}
    \end{center}
\end{table*}
\begin{table}[]
    \centering
    \caption{ Classification results on SEED-V dataset
}
    \label{tab:eleven}
    \begin{tabular}{|c|c|c|}
    \hline
        \textbf{Year} & \textbf{Model} & \textbf{Accuracy} \\ \hline
        2019 & cBEGAN \cite{luo2019gan} & 68.32\% \\ \hline
        2019 & CNN and LSTM \cite{guo2019multimodal} & 79.63\% \\ \hline
        2020 &  DGCCA \cite{lan2020multimodal} & 82.11\% \\ \hline
        2020 & SVM \cite{wu2022investigating}  & 84.51\% \\ \hline
        2021 &  DCCA \cite{liu2021comparing} & 85.3\% \\ \hline
         \textbf{2023} & \textbf{Proposed model} & \textbf{99.80\%} \\ \hline
    \end{tabular}
\end{table}
Apart from this, we evaluate the SEED-V dataset with individual modality. Table \ref{tab:ten} illustrates the comparison between single-modality and multi-modality. With the single eye movement feature, we obtain 61\% testing accuracy whilst the power spectrum and DE features exhibit 73\%, and 26\% testing accuracy respectively. We obtain effective performance when we combine all the features. On top of that, to the best of our exploration, our method has prevailed the best performance among existing systems on the SEED-V dataset which is given in Table \ref{tab:eleven}. In addition, to evaluate the proposed model more accurately we generate a confusion matrix report and examine with precision, recall, and f1-score that are shown in Table \ref{tab:twelve} and Fig. \ref{fig:cm}. respectively.

\begin{table}[]
    \centering
    \caption{Performance report for our proposed 1D-CNN on SEED-V dataset}
    \label{tab:twelve}
    \begin{tabular}{|p{1.4cm}|p{0.6cm}|p{0.35cm}|p{0.35cm}|p{0.5cm}|}
    \hline
    \textbf{Class-Label} & \textbf{Precision} & \textbf{Recall} & \textbf{F1-score} & \textbf{Data Points} \\ \hline
       Disgust - 0  & 1.00 & 1.00 & 1.00 &  118,542 \\ \hline
       Fear - 1  & 1.00 & 1.00 & 1.00 & 144,408 \\ \hline
       Sad - 2 & 1.00 & 1.00 & 1.00 & 183,908 \\ \hline
       Neutral - 3 & 1.00 &  1.00 & 1.00 &  141,745
       \\ \hline
       Happy - 4 & 1.00 & 1.00 & 1.00 & 117,213 \\ \hline
    \end{tabular}
    
\end{table}
\section{Conclusion}
In this paper, we propose a subject-dependent end-to-end emotion recognition system using the 1D-CNN model which enhances the classification performance. We decompose raw EEG signals into frequency bands utilizing FFT and extract the power spectrum feature which provides high-resolution information. We combine three features including the power spectrum, eye movement, and differential entropy. Afterward, we apply these features in 1D-CNN with two convolution layers. The classification result accounts for 99.80\% which outperforms the performance of existing models utilizing the SEED-V dataset. However, a limitation of this research is to sole use of the SEED-V dataset whilst there are significant numbers of publicly available datasets for emotion recognition. Hence, along with the SEED-V dataset, we will explore other datasets to enhance the generalization reliability. Furthermore, in terms of splitting the dataset into Train-Test, we will expand our study in the future to propose a subject-independent end-to-end beneficial emotion recognition system. \newline
\textbf{Data and code availability:} Data for this study is available at \href{https://bcmi.sjtu.edu.cn/home/seed/seed-v.html}{https://bcmi.sjtu.edu.cn/home/seed/seed-v.html}. Code will be publicly available on the GitHub repository, and this will be added after the review process when author names are disclosed.

\end{document}